\documentclass[aip,
rsi,amsmath,amssymb,
reprint,
final,
author-year,author-numerical,%
]{revtex4-1}

\usepackage{graphicx}
\usepackage{dcolumn}
\usepackage{bm}
\usepackage{natbib}
\usepackage{verbatim}
\usepackage{float}

\begin{document}
\title[]{High resolution ion trap time-of-flight mass spectrometer for cold trapped ion experiments}

\author{P. C. Schmid}
\email{philipp.schmid@jila.colorado.edu}
\author{J. Greenberg}
\author{M. I. Miller}
\author{K. Loeffler}
\author{H. J. Lewandowski}
 \affiliation{JILA and the Department of Physics, University of Colorado, Boulder, Colorado 80309-0440, USA.}

\date{\today}

\begin{abstract}
Trapping molecular ions that have been sympathetically cooled with laser-cooled atomic ions is a useful platform for exploring cold ion chemistry. We designed and characterized a new experimental apparatus for probing chemical reaction dynamics between molecular cations and neutral radicals at temperatures below 1 K. The ions are trapped in a linear quadrupole radio-frequency trap and sympathetically cooled by co-trapped, laser-cooled, atomic ions. The ion trap is coupled to a time-of-flight mass spectrometer to readily identify product ion species, as well as to accurately determine trapped ion numbers. We discuss, and present in detail, the design of this ion trap time-of-flight mass spectrometer, as well as the electronics required for driving the trap and mass spectrometer. Furthermore, we measure the performance of this system, which yields mass resolutions of $m/\Delta{}m \geq 1100$ over a wide mass range, and discuss its relevance for future measurements in chemical reaction kinetics and dynamics.
\end{abstract}

                           
\maketitle

\section{\label{sec:introduction}Introduction}
Reactions of molecular cations with neutral molecules and radicals are important in many environments, like the interstellar medium.\cite{Larsson2012,Tielens2013} Detailed understanding of these reactions would benefit from experiments at low temperatures, where the internal quantum states and external motion can be controlled with high precision. Preparing the reactants in a limited distribution of quantum states and at cold temperatures can allow for studies of how the initial states determine reaction pathways.\cite{Heazlewood2015,Willitsch2012} However, experimental measurements of reaction rates and chemical dynamics between cations and neutral molecules, especially radicals, at low temperatures under controlled conditions are challenging. This is a direct consequence of the low densities of ions and radicals typically achieved in the lab, as well as lack of molecular purity in radical samples. To overcome the hurdle of low density samples, recent experiments have employed the use of ion traps to dramatically extend the interaction times over molecular beam experiments. Thus, cold ion-radical reaction measurements become possible.

 One experimental approach to create cold, trapped molecular ions is to laser cool atomic ions confined in a linear Paul trap \cite{Eschner2003} and use them to sympathetically cool co-trapped molecular ions.\cite{Baba2002a,Ostendorf2006} In these systems, the translational temperatures of the ions are small compared to the Coulomb repulsion, and so-called Coulomb crystals are formed. \cite{Drewsen2015} Although the number density of the ions in a Coulomb crystal is space-charge limited to 10$^7$ \textendash ~10$^9$ ions cm$^{-3}$, the ionic reactants can be trapped for long periods of time, allowing for precise measurements of reactions even with such low density samples. Using these trapped Coulomb crystals, several groups have studied chemical reactions between atomic ions and neutral molecules, including, but not limited to, Mg$^+$+H$_2$,\cite{Molhave2000} Be$^+$+HD,\cite{Roth2006} Ca$^+$+CH$_3$F,\cite{Willitsch2008} Yb$^+$+Ca,\cite{Rellergert2011} Ba$^+$+Rb, \cite{Hall2013} and Ca$^+$+ND$_3$.\cite{Okada2013} Other experiments extended such reaction studies to sympathetically cooled molecular ions and neutral molecules, including H$_3$O$^+$+NH$_3$,\cite{Baba2002} CaO$^+$+CO, \cite{Drewsen2002} N$_2^+$+H$_2$, \cite{Roth2006a} OCS$^+$+NH$_3$,\cite{Bell2009a} N$_2^+$+N$_2$,\cite{Tong2010} N$_2$H$^+$+CH$_3$CN,\cite{Okada2013} and reactions involving conformer-selected neutral molecules. \cite{Roesch2014}
 
An important step in all of these reaction experiments is the detection of reaction products. It is possible to identify probable reaction products in trapped ion experiments by changes in the Coulomb crystal structure\cite{Roth2006,Bell2009a} or by secular excitation of the trapped ions.\cite{Baba2002b,Drewsen2004,Roth2007} But these methods lack the required sensitivity for experiments on complex reactions with multiple reaction pathways. Coupling of a time-of-flight mass spectrometer (TOF-MS) to the ion trap can boost the detection sensitivity \textemdash ~ even down to the single ion level \textemdash ~ while allowing for accurate determination of the number and charge-to-mass ratio of all trapped ions in a single recorded spectrum. 

Time-of-flight mass spectrometers can be easily coupled to linear Paul traps along the axial direction, but the mass resolution is limited by the extended size of the trapped ion cloud along the extraction axis. \cite{Seck2014} This limitation can be overcome, and higher mass resolutions can be reached, at the cost of complex ion optics.\cite{Collings2001} Recently, several experiments \cite{Schneider2014,Meyer2015,Deb2015,Roesch2016} have demonstrated extraction of the ions along one of the radial trap axes into the TOF-MS, where the trap electrodes also act as DC repeller plates of the TOF-MS. Using this technique, both Schneider et al. \cite{Schneider2014}  and Meyer et al. \cite{Meyer2015} achieve a mass resolution of m/$\Delta$m $\leq$ 500. In contrast, R\"osch et al. \cite{Roesch2016} installed dedicated TOF-MS repeller plates, mounted in between the ion trap rods to enhance their detection efficiency.  While this helped to increase the mass resolution to about m/$\Delta$m $\approx$ 700, this specific electrode design distorts the RF trapping field, requiring compensation potentials for efficient ion trapping. Also, the increased mass resolution is limited to a single, individual mass peak at any one time.

Here, we report the development of a new linear ion trap TOF-MS experiment dedicated to the measurement of ion-radical reaction dynamics. While the general layout is similar to previous published apparatuses, \cite{Schneider2016,Roesch2016} our set-up has the advantage of having significantly simpler trap and TOF-MS electronics and an increased mass resolution over a large range of masses simultaneously.

In this paper, we begin by discussing the design of the combined linear ion trap time-of-flight mass spectrometer (LIT TOF-MS)  (Sec. \ref{sec:instrument}). Next, we present the driving electronics for operating the LIT TOF-MS system (Sec. \ref{sec:instrument.electronics}) In Sec. \ref{sec:results}, we show the results for the optimized system, demonstrating the advantages of our design. We conclude in Sec. \ref{sec:summary} by giving an outlook for future experiments planned with ion trap TOF-MS apparatus.
\section{\label{sec:instrument}Experimental Design}
Our experimental apparatus consists of a segmented linear quadrupole ion trap and a linear TOF-MS, coupled radially to the ion trap. We trap $^{40}$Ca$^+$ and perform laser cooling to form Coulomb crystals with temperatures $<$ 1 K. Molecular ions can be co-trapped and sympathetically cooled by the atomic ions. To facilitate high cooling rates and have Coulomb crystal lifetimes of several hours, the vacuum chamber is maintained at a base pressure of $<$5$\times$10$^{-10}$ torr using a 600 l/s turbo pump (Pfeiffer HiPace 700) backed by a dry roughing pump (Adixen ACP-15). A titanium sublimation pump is used in addition for enhanced pumping of certain residual reactive contaminants. Finally, to eliminate background water in the chamber, a cold finger filled with liquid nitrogen is used.

Trapped ions are detected either by the fluorescence of the atomic ions or by ejecting all trapped ions into the TOF-MS. To detect mass spectra with high mass resolution over a wide mass range, the trapping RF amplitude must be switched off within a couple RF periods before the HV extraction pulses are applied to the trap electrodes. Here dedicated RF drive electronics were built and characterized. The next sections describe in detail these individual components of the apparatus.

\subsection{\label{sec:instrument.setup}Ion trap time-of-flight set-up}

\subsubsection{\label{sec:instrument.trap}Ion trap and TOF-MS combination}

\begin{figure}[h]
\includegraphics[width=\columnwidth]{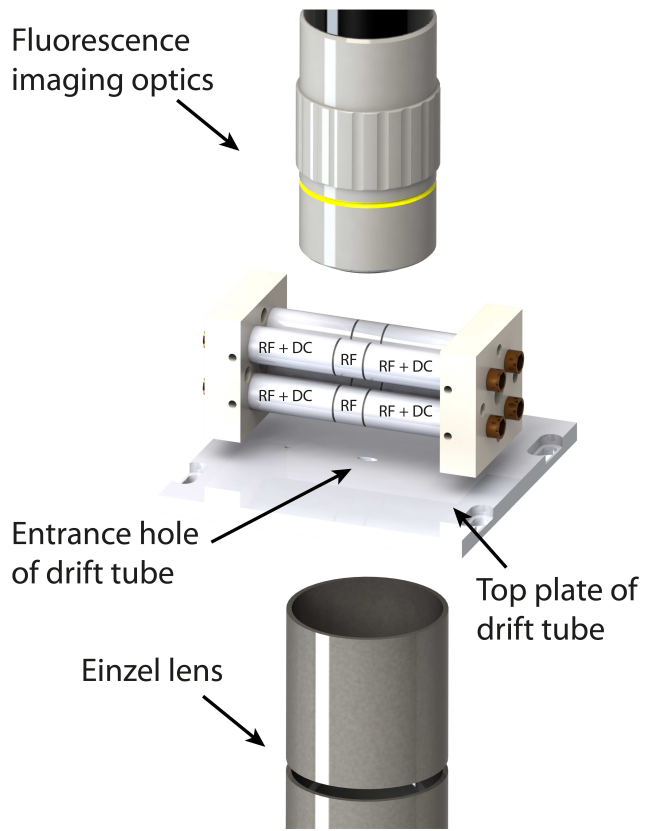}%
\caption{\label{fig:fig1}  Schematic of the ion trap and detection optics. The ion trap has four rods, which are broken into three segments each. RF potentials are applied to all 12 segments. In addition, small DC potentials are applied to form a trapping field along the longitudinal axis.  Directly above the ion trap sits a 10x microscope objective to image the fluorescence from the Ca ions onto a EMCCD camera. The drift tube of the TOF-MS is located directly below the ion trap and includes an Einzel lens (deflection plates not shown).
}
\end{figure}

The central component of the experimental apparatus is a segmented, linear quadrupole ion trap built from four rods (Fig. \ref{fig:fig1}). RF potentials of several hundred volts peak-to-peak are applied to the rods as shown in Fig. \ref{fig:fig1} to produce a time-averaged trapping potential in the radial dimension.  Additionally, each rod is divided into three electrically isolated segments, which allow for the application of individual DC voltages on each segment to confine the ions along the longitudinal axis. The DC voltages can also be used to make small adjustments to the position of the ion cloud in three dimensions to center the ions on the node of the RF potential, thus minimizing RF heating of the ions.

Our trap has an inscribed radius of $r_0$ = 3.91 mm  and a trap rod radius of r = 4.5 mm. This ratio produces a nearly perfect radial quadrupole field in the center of the trap, \cite{Denison1971} reducing the micromotion heating of the ions. The central segments of the ion trap have a length of 2$\times$z$_0$ = 7 mm, while the outer segments have a length of z = 20 mm. These lengths give rise to an approximately harmonic axial potential in the center of the trap. To trap the laser-cooled atomic ions, the RF frequency applied to the electrodes is $\Omega_{RF}$ = 2$\pi\times$3.1 MHz with an amplitude up to V = 250 V. Axial confinement is achieved with DC potentials of $V_{DC} \leq$ 10 V applied to the outer segments of the rods (Fig. \ref{fig:fig1}).
To produce atomic calcium atoms for loading into the trap, we use an effusive calcium beam generated by a home-built, resistively heated Ca oven. The oven is made out of a thin-walled stainless steel tube, crimped on both ends, with an 1 mm$^2$ opening at one end. The oven is mounted 300 mm from the trap center and uses a collimating aperture in the beam path to minimize Ca deposition on the trap rods.  A simple beam shutter is placed in the atomic beam path to block the Ca impinging on the trap after the loading process. This allows the Ca oven to be kept at a consistent temperature at all times, while reducing the influence of the oven products on the trapped ions. The atomic beam of $^{40}$Ca is then ionized in the trap center using non-resonant photo-ionization at 355 nm (Continuum Minilite II, 10 Hz, 5 ns pulse length, up to 7 mJ/pulse). In addition to the Ca oven, we installed a set of three alkali atom sources dispensers (SAES) \textemdash ~ K, Rb and Cs \textemdash ~  for calibration of the TOF-MS spectra.

The trapped atomic ions are laser cooled on the 4s$^2S_{\frac{1}{2}}$- 4p$^2P_{\frac{1}{2}}$ transition of $^{40}$Ca$^+$ using a fiber-coupled 397 nm diode laser (NewFocus, 3.5 mW, 600 $\mu m$ beam diameter). The ions are repumped from a dark state on the 4p$^2P_{\frac{1}{2}}$ - 3d$^2D_{\frac{3}{2}}$ transition using a second fiber-coupled diode laser at 866 nm (NewFocus, 9 mW, 2 mm beam diameter). The two diode laser frequencies are measured by a wavemeter (High Finnesse/{\AA}ngstrom WSU-30), which is calibrated daily to an additional diode laser at 780 nm locked via saturated absorption spectroscopy to a transition in $^{87}$Rb. The calcium cooling laser frequencies are locked via a slow LabVIEW controlled servo based on the wavemeter readings. For efficient cooling of ion crystals down to the single ion level, the 397 nm laser beam is directed into the center of the trap from three orthogonal directions (two counter-propagating beams along the axial dimension and one beam along the radial dimension). The axial beams are of equal intensity, while the radial beam has about 2 $\%$ of the total power. The radial beam is required only to efficiently cool small strings of ions, but is unnecessary for large crystals. The emitted fluorescence at 397 nm from $^{40}$Ca$^+$ is collected by a 10$\times$ microscope objective (Mitutoyo) and imaged onto an EMCCD camera (Andor iXon 897). This allows real-time, non-destructive detection of ions stored in the trap as shown in Fig. \ref{fig:fig4}.

In addition to imaging the fluorescence to determine the number of Ca ions, we have incorporated a radially extracted TOF-MS to the ion trap to greatly increase the accuracy of measurements of non-fluorescing ions. The TOF-MS consists of two stages of longitudinal electric fields that extract the ions from the trap and accelerate them into a longitudinal zero potential drift tube. The entire length of the TOF-MS is 540 mm. The trap electrodes themselves create the extraction and acceleration electric fields for the TOF-MS. This is accomplished by applying large DC potentials to the four trap rods, where the top two are at one positive potential and the bottom two trap rods are at another, lower, positive potential. A drift tube, held at ground potential, is placed between the ion trap and the ion detector to minimize the influence of stray fields on the post-accelerated ions. The entrance into the drift tube, which is 22 mm from the center of the trap, is limited by a 2 $\times$ 5 mm elliptical hole located directly below the center of the trap. Using a small opening reduces the fringe fields from the trap electrodes penetrating into the drift tube. Directly after the entrance of the drift tube are a set of four deflection plates for beam steering and a cylindrical Einzel lens for beam focusing. At the end of the drift tube, a 40 mm diameter micro-channel plate (MCP) detector (double-stack in Chevron configuration, Jordan-TOF) produces a TOF spectrum from the ions.

During an experimental run, the process to detect the ions is as follows. First, the RF trapping field is rapidly quenched over $\sim$2 RF cycles. Second, after an optimized time delay, high voltage is pulsed onto the four trap rods. This creates a two-stage acceleration field in a Wiley-McLaren configuration,\cite{Wiley1955} increasing the mass resolution of the recorded spectra. Third, the ions pass through an Einzel lens, whose potentials have been set to maximize ion transmission as well as mass resolution. Finally, the ions hit the MCP detector creating a current that is split by a 50/50 RF power splitter (MiniCircuits ZSC-2-1). The currents are converted to voltages using two separate channels of a fast digital storage oscilloscope. The two channels of the oscilloscope are set with different vertical scaling to allow for the dynamic range required to simultaneously measure the large number of calcium ions (100 \textendash~1000) on one channel, while still having the sensitivity on the other channel to measure lower numbers of (molecular) product ions (1 \textendash~100).

\subsection{\label{sec:instrument.electronics}Electronics for driving the linear ion trap TOF-MS}

The electronics for operating a combined ion trap/TOF-MS instrument need to perform three distinct operations : (1) application of RF and small DC fields for ion trapping, (2) rapid quenching of the RF voltage to create zero electric fields to prepare for ion extraction, and (3) rapid application of high DC voltages for ion extraction. First, during the normal trapping mode, the electronics must produce two RF sine waves at the same frequency (a few MHz), but with a 180$^\circ$ phase shift between the two signals. The amplitude of these signals must be variable up to 250 V.  Second, for optimal extraction of the ions into the TOF-MS, the RF fields must be turned off within 1-2 RF cycles. Otherwise, the RF fields will distort the extraction field, reducing the mass resolution of the TOF-MS. Finally, to direct the ions into the drift tube with well defined starting conditions, two HV pulses with fast rise times and different amplitudes must be applied to the four trap rods to extract and accelerate the ions towards the MCP.

A varitey of different RF drive electronics have been described by other groups using a linear ion trap TOF-MS. R\"osch et al.,\cite{Roesch2016} uses separate ion trap and ion extraction electrodes, allowing them to use commercial RF and HV electronics. The group at Oxford implemented a so-called ``digital trap,'' which uses square-wave RF potentials.\cite{Deb2015} Because these specific driving electronics do not use a resonant tank circut, the potentials can be turned off almost instantaneously. Unfortunately, the digital trapping configuration leads to a very flat trapping potential, limiting its usefulness for creating cold Coulumb crystals. The same group also demonstrated a sinusoidally driven ion trap. \cite{Meyer2015} With this configuration, the RF voltage is turned off in about 1.5 $\mu s$ ($\sim$ 4-5 RF cycles) before HV extraction potentials are applied with an amplitude of less than 400 V and rise times of several hundred ns. Schneider at al. \cite{Schneider2016} developed dedicated drive electronics for their ion trap TOF-MS experiment. They demonstrated fast quenching of the trapping RF within a single RF cycle and application of HV pulses with amplitudes up to 1380 V and a rise time of $\sim$250 ns. Because of the design of their circuit, an arbitrary time delay between turning off of the RF and activation of the HV pulse is not possible, and thus the Wiley-McLaren TOF-MS condition \cite{Wiley1955} is not met. Additionally, for each individual trap electrode a full drive circuit \textemdash ~  e.g. RF drive, RF quench and HV pulse circuit \textemdash ~  is required, which requires advanced timing-synchronization between all 12 RF drive circuits.

\begin{figure*}
\centering
\includegraphics[width=\textwidth]{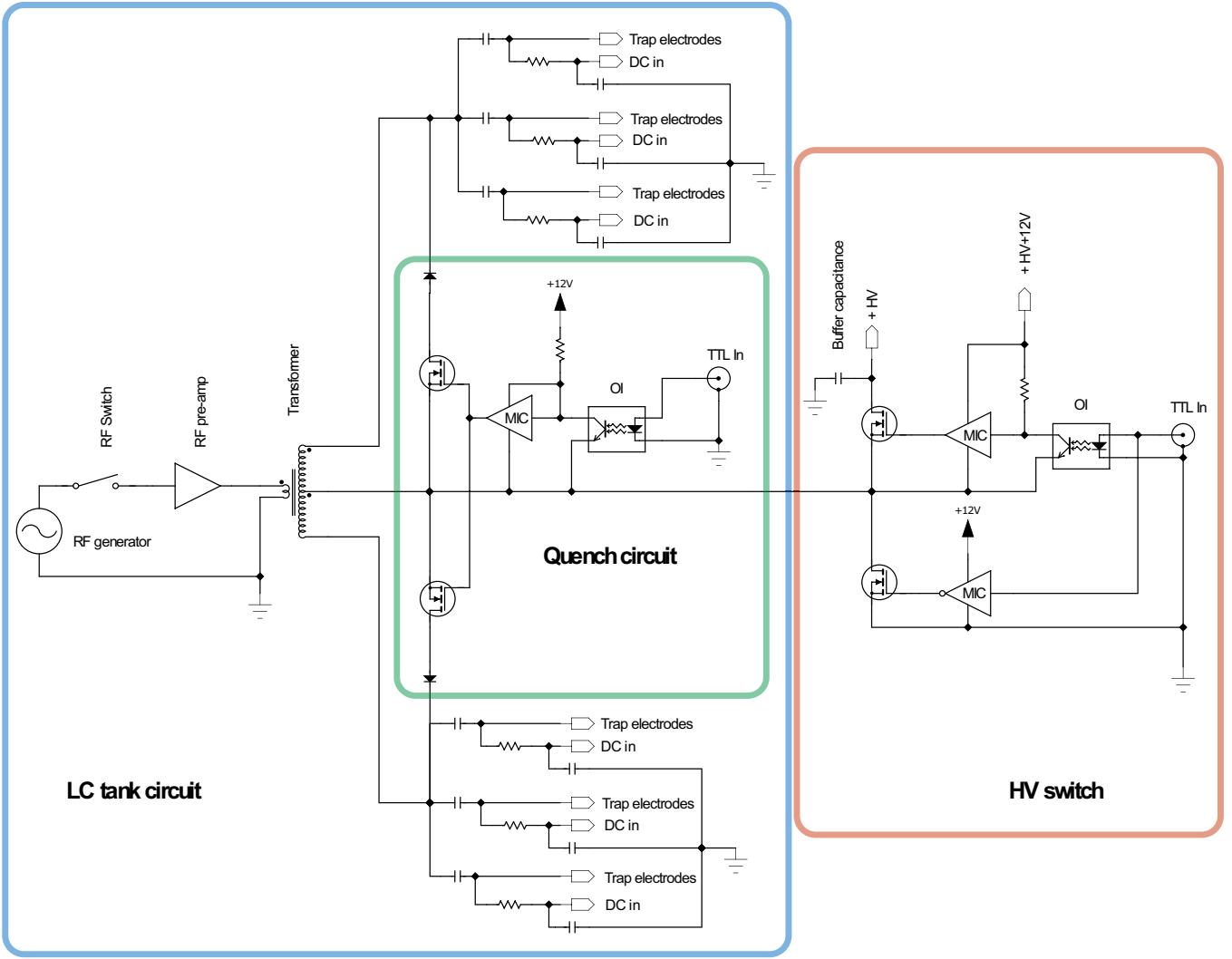}
\caption{A schematic diagram of the RF drive electronics. The diagram shows the three different parts (marked by colored boxes) required for ion trapping and extraction into the TOF-MS (only single side of bifilar winding is shown). Two RF signals, separated in phase by 180$^\circ$, are produced by an LC tank circuit (blue box), where the inductance comes from a bifilar wound transformer coil and the capacitance is from a combination of the capacitance of the SHV cables and trap electrodes. To quickly quench the RF amplitude, the high-sides of the coil can be short-circuited to the center tap by a diode-MOSFET chain (green box). The HV pulses with a fast rise time ($<$300 ns) and variable amplitude are applied by two power MOSFETs in a push-pull configuration (red box).}
\label{fig:fig2}
\end{figure*}

In the following, we present a simplified version of an electronic circuit to drive an ion trap TOF-MS with the schematic shown in Fig. \ref{fig:fig2}. The electronics are capable of generating two RF potentials 180$^\circ$ out of phase with an amplitude of several hundred volts in the 1-5 MHz range. The RF amplitude can be quenched within 1-2 RF cycles and two fast HV pulses of different amplitudes up to 2.5 kV can be produced. Unlike other published implementations of such a combined system, our design can have an arbitrary time delay between the RF quench and the HV pulses. This preserves the Wiley-McLaren condition \cite{Wiley1955} for the TOF-MS.

Our design can be divided roughly into three distinct sections: a LC tank circuit for generating the RF trapping potentials, a quench circuit for rapidly switching off the RF, and an HV switch for creating the extraction pulses for the TOF-MS. In the following sections, the individual parts of the circuit are described in more detail.

\subsubsection{\label{sec:instrument.electronics.RFdriver}LC tank circuit}

For driving the ion trap with the required RF field, we use a LC tank circuit, as shown in Fig. \ref{fig:fig2}. The inductor, L, is a homemade transformer coil, while the ion trap electrodes and SHV cables form the capacitance of the circuit. Low voltage RF is generated by a signal generator (Stanford Research Systems, DS345) and amplified by a home-built RF pre-amplifier before driving the primary side of a toroidal transformer core (Mircometal, T200-2). For the secondary side, we use a center-tapped bifilar winding, which is required for the application of the two individual HV pulses for ion extraction. The center-tap creates two RF signals with 180$^\circ$ phase shift. During ion trapping operation, the center-tap is kept at ground potential and pulsed to HV for mass analysis with the TOF-MS. Each electrode is also AC coupled by individual blocking capacitors. By choosing the appropriate values of L and C, the RF frequency is set to $\Omega \approx $2$\pi\times$3.1 MHz. To apply individual DC voltages on each ion trap segment, a low-pass filter couples the DC to the RF on the trap side of the blocking capacitors.

\subsubsection{\label{sec:instument:electronics.Quench}Quench circuit}

As mentioned previously, fast quenching of the RF before ion extraction is required for high resolution TOF-MS mass spectra. If the RF is turned off by removing the input signal only, a long ring-down period of several tens of microseconds will be observed due to the high quality factor of the LC tank circuit. To reduce the turn-off time to within 1-2 RF cycles, we implement an active damping circuit (quench) into our LC tank circuit (Fig. \ref{fig:fig2}). The high sides of each secondary winding are connected to the respective center-tap by a diode-MOSFET chain. When the MOSFET closes it creates a low-resistance current path, effectively shorting the coil. This stops the RF output in a minimal time. Quenching is implemented on the secondary winding of each side of the center tap simultaneously to reduce the quench time by nearly a factor of two. Additionally, the input RF is turned off via an RF switch (Mini-Circuits ZYSW-2-50DR) on the primary side of the coil at the same time as the quench. This further reduces the turn-off time of the RF potentials on the trapping electrodes.
\begin{figure}[h]
\centering
\includegraphics[width=\columnwidth]{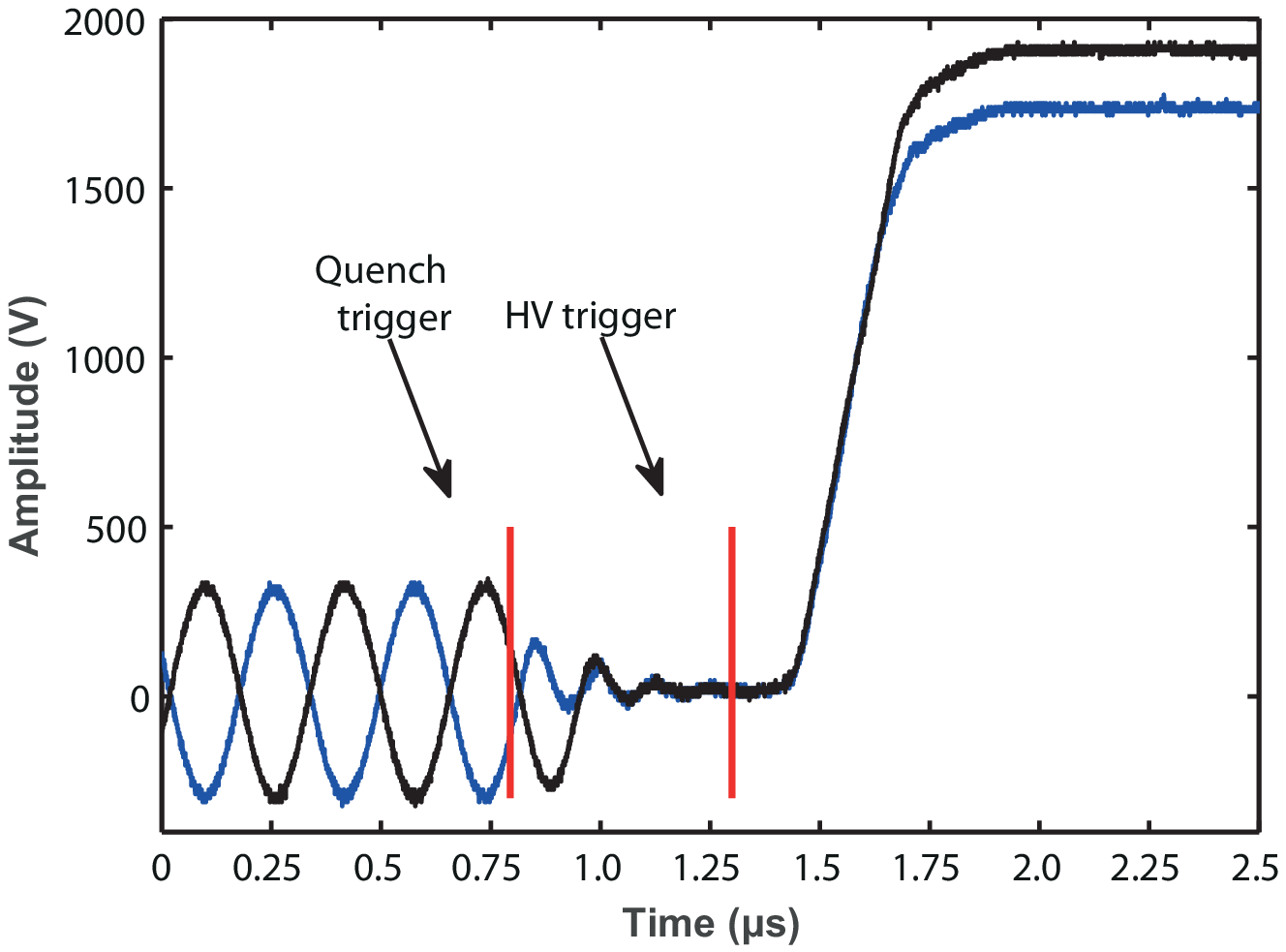}
\caption{Measured potentials on one top and one bottom trap electrode during trapping, RF quenching, and ion extraction. The trapping RF waveform is shown between 0 and $\sim$ 0.75 $\mu$s. After the RF quench is activated (marked by the first vertical line) the RF amplitude drops to less than $<$10\% of its initial value within 1 \textendash 2 RF cycles. At an arbitrary time delay afterwards, the HV pulse is activated (indicated by the second vertical line). The rising edge shows a smooth increase of voltage with a 10\% - 90\% rise time of 220 ns.}
\label{fig:fig3}
\end{figure}
The time of arrival of ions on the MCP depend sensitively on the time during the RF cycle when the quench is triggered, thus we trigger the quench circuit at a zero crossing of the RF signal. This ensures that the ions always experience similar electric field distributions, reducing RF phase-dependent jitter in the mass spectra.

Additionally, because the HV pulses are applied on the center-taps, the quench circuit must be able to float up to 2.5 kV together with the coil during ion extraction. To accomplish this, the MOSFETs (MTP2N50E) are controlled by a driver chip (MIC4429, MIC in Fig. \ref{fig:fig2}) triggered by an optocoupler (ACNV4506, OI in Fig \ref{fig:fig2}), where the entire circuit is powered by a DC/DC converter (MIR512). This configuration allows the circuit to float to the required HV during the extraction.

\subsubsection{\label{sec:instrument:electronics:HVswitch}HV switch}

After the RF signal is quenched, a fast rise time HV pulse is applied to extract the ions out of the trap and accelerate them into the drift tube. To produce the HV pulse (up to 2.5 kV), we use a home-built HV switch, shown in Fig. \ref{fig:fig2}. The central elements are two MOSFETs (IYXS IXTL2N450) in a push-pull configuration. The low-voltage side MOSFET is driven by a single MOSFET gate driver (MIC4429, MIC in Fig. \ref{fig:fig2}) and the gate of the high-voltage side MOFSET is driven by a opto-coupled gate driver(MIC4420,MIC in Fig. \ref{fig:fig2}), which is powered by a DC/DC converter. This design is similar to the quench circuit. 

Two of these HV switches are used to apply different potentials to the top two and bottom two trap rods. The two switches are connected directly to the center-tap of each of the bifilar windings. During the RF trapping configuration, the high-voltage side MOSFET is open while the low-voltage side MOSFET is closed, thus providing a ground connection for the center-tap of the secondary windings.

The output of the combined quench and HV switch circuits is shown in Fig. \ref{fig:fig3}. Here, the two phases of the RF are measured by attaching a pair of 100$\times$ oscilloscope probes to two trap electrodes. The first vertical red line indicates the MOSFET trigger to short the coil and switch off the input signal. The RF amplitude then decays within two RF cycles. We measure a 100\% to 10\% decay time of 360 ns, which is nearly independent of the initial RF amplitude. We adjust the time between the quench and HV pulse triggers to optimize the detection efficiency and mass resolution of our TOF-MS. The second vertical red line indicates the HV switch trigger. The delay of the HV switch with respect to the quench trigger was set to 300 ns, which results in an optimized mass resolution. The HV pulses on the top and bottom trap electrodes have identical turn-on characteristics and reach the design amplitude of 2 and 1.84 kV at the same time (the mismatch between objective and measured amplitudes is due to the oscilloscope probe capacitance). The 10\% to 90\% rise time of the HV pulses is measured to be 220 ns and displays a smooth increase in voltage. The rapid quench and HV pulses coupled with a controllable delay between the the two operations leads to a high mass resolution in the TOF-MS spectra.

\section{\label{sec:results}Results}
\subsubsection{\label{sec:results:general}Ion trap and TOF-MS performance}

\begin{figure}[h]
\centering
\includegraphics[width=\columnwidth]{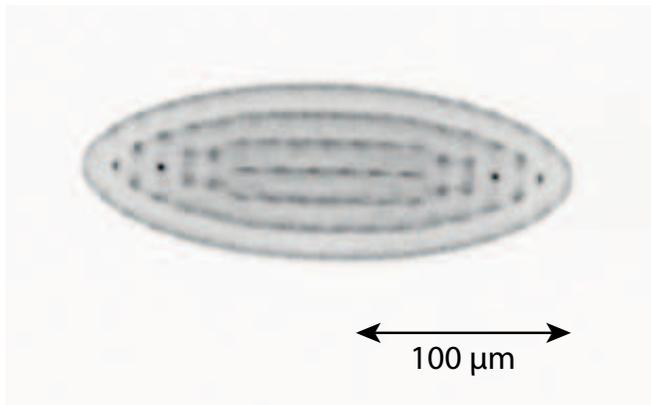}
\caption{A typical fluorescence image of a Coulomb crystal in the ion trap. This crystal contains about 370 ions. The image integration time was 0.2 sec. }
\label{fig:fig4}
\end{figure}

An experimental run begins by creating Ca ions. They are created in the center of the trap via non-resonant photo-ionization using a frequency tripled Nd:YAG laser at 355 nm. We can create Coulomb crystals that contain between one ion to several thousand ions by varying the 355 nm intensity, the Ca oven temperature, and the number of photo-ionization pulses. To create a Ca crystal with 700 ions, we typically use 100 pulses of 5 mJ/pulse light focused via a 200 mm focal length lens. To minimize the time it takes to load ions into the trap and cool them to below the crystallization temperature, we load the Ca ions at low RF amplitudes. Next, the RF amplitude is ramped to its final value, pushing the ions into the center of the trap, increasing the overlap with the cooling lasers, and therefore the cooling rate. Figure \ref{fig:fig4} shows a typical Coulomb crystal after loading. The number of ions in this crystal was determined to be 370 from the corresponding TOF-MS spectrum.

\begin{figure}[h]
\centering
\includegraphics[width=\columnwidth]{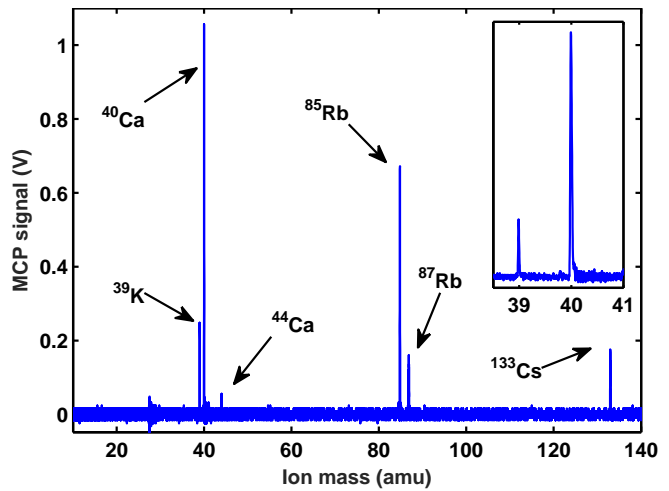}
\caption{A single-shot TOF-MS mass spectrum of trapped, laser-cooled $^{40}$Ca$^+$ and co-trapped $^{39}$K$^+$, $^{44}$Ca$^+$, $^{85,87}$Rb$^+$ and $^{133}$Cs$^+$. This Coulomb crystal contained 698 ions with 267 of them being $^{40}$Ca$^+$. Apart from these ions, no further mass peaks can be identified. The inset shows the resolving power of the TOF-MS by the clear separation of the $^{39}$K$^+$ and $^{40}$Ca$^+$ peak. All peaks in this spectrum show a mass resolution greater than 1000.}
\label{fig:fig5}
\end{figure}

In addition to loading Ca ions into the trap to be laser cooled, we can load in other atomic (or molecular) ions where they will be sympathetically cooled by interactions with the Ca ions. Using SAES atom dispensers as the source of neutral atoms, we were able to ionize, co-trap, and cool various alkali atoms in our trap, including $^{39}$K$^+$, $^{85,87}$Rb$^+$, and $^{133}$Cs$^+$. These ions serve as a convenient source of different mass ions to be able to calibrate our mass spectrometer.  Figure \ref{fig:fig5} shows a typical single shot (non-averaged) mass spectrum of a multi-component Coulomb crystal. Because the mass range of interest is centered around 40 amu, we optimized the TOF-MS voltages and timings to this region.

The mass spectrum in Fig. \ref{fig:fig5} shows several distinct ion peaks, which can be directly assigned to the ionized atoms. The peak with the strongest intensity is created by $^{40}$Ca$^+$. The smaller peak to the left is $^{39}$K$^+$, while the very small peak on the right hand side of $^{40}$Ca$^+$ is another isotope of calcium, $^{44}$Ca$^+$. The relative peak heights of $^{40}$Ca$^+$ and $^{44}$Ca$^+$ represent the natural isotopic abundance.  At higher masses, $^{85}$Rb$^+$, $^{87}$Rb$^+$, and $^{133}$Cs$^+$ can be identified. Apart from these masses, no further ions are present. The small feature around mass 30 amu is due to electrical noise in the read-out electronics and does not represent a real ion signal. The inset of Fig. \ref{fig:fig5} shows a zoom of the $^{39}$K$^+$ and $^{40}$Ca$^+$ peaks. These two neighboring masses are clearly separated, with the peaks showing no identifiable shoulders or peak splitting.

\begin{figure}[h]
\centering
\includegraphics[width=\columnwidth]{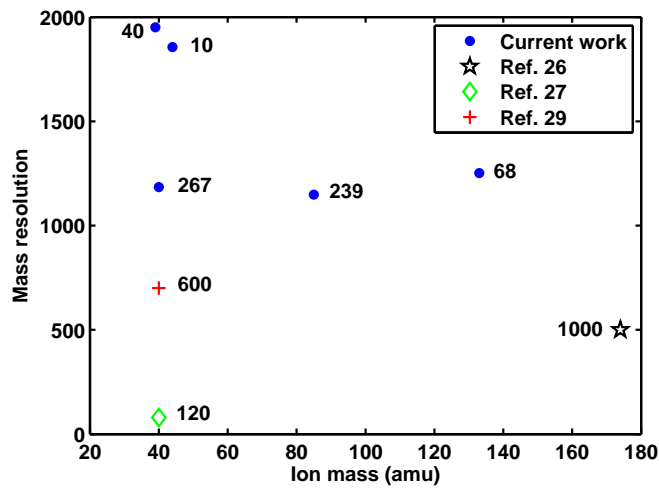}
\caption{Measured mass resolutions for various ion trap TOF-MS experiments including the current work. This plot shows the measured mass resolution of several different detected mass peaks ($^{39}$K$^+$, $^{40}$Ca$^+$, $^{44}$Ca$^+$, $^{85}$Rb$^+$ and $^{133}$Cs$^+$) from Fig. \ref{fig:fig5}. The number next to the point represents the number of ions at that mass. While we observe the influence of the ion mass and number on the resolution, a high mass resolution can be achieved simultaneously over a wide mass range within a single shot.}
\label{fig:fig6}
\end{figure}

\begin{figure}[h]
\centering
\includegraphics[width=\columnwidth]{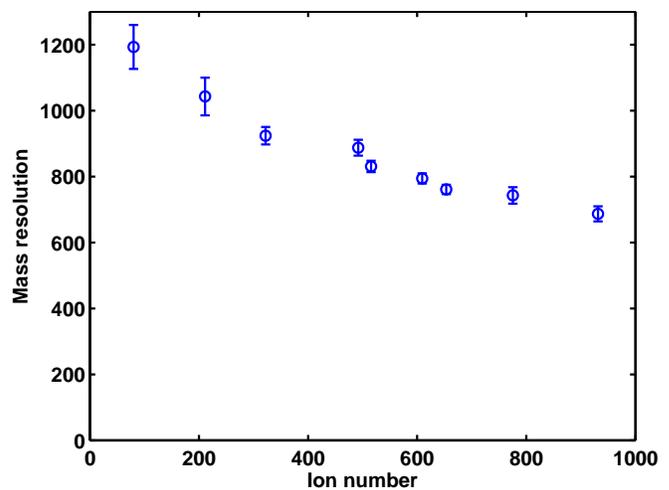}
\caption{Dependence of mass resolution on trapped ion numbers. Pure $^{40}$Ca$^+$ Coulomb crystals of different sizes are trapped and detected in the TOF-MS. We observe a decrease of mass resolution with increasing ion number.}
\label{fig:fig7}
\end{figure}

We determine the mass resolution of our TOF-MS and how it depends on the mass of the ion being detected and on the number of ions in the trap. The mass resolution is defined as R = t/(2$\Delta$t) = m/$\Delta$m. Using this definition, every peak in the mass spectrum in Fig. \ref{fig:fig5} has a mass resolution of greater than R$>$1200. The mass resolution is expected to decrease with increasing ion mass and ion numbers. To investigate this relationship in our system, we plot the measured mass resolution as a function of mass (Fig. \ref{fig:fig6}), where the number adjacent to each point represents the total number of ions at this mass in the trap. The data presented for the current work are taken from the spectrum shown in Fig. \ref{fig:fig5}. Our system was optimized for the maximum resolution of m/z = 40 amu. In a single shot, however, we observe mass resolutions above 1200 for masses even $>$ 3 times the optimized mass. This high mass resolution across a wide range of masses in a single spectrum is critical for reaction experiments where the product masses can be very different than calcium. The \textit{total} ion number of ions in the Coulomb crystal has no observable influence on the observed mass resolution of each observed mass channel, as shown in Fig. \ref{fig:fig6}. The reduction of mass resolution due to the number of ions within a single mass channel are investigated in a different experiment and is discussed below (Fig. \ref{fig:fig7}).
We can compare our results to other published results of mass resolutions from radially extracted LIT TOF-MS. As shown in Fig. \ref{fig:fig6}, the resolution from the work of R\o"sch et al. \cite{Roesch2016} is slightly lower than the resolution we obtain, but because of the complexity of their system, they can realize this resolution for only one mass at a time. This is in contrast to our set-up, where high mass resolution is possible for all masses within a mass window greater than 100 amu in a single shot. The UCLA group also measures a slightly lower resolution, but their published number is for a larger crystal and a heavier mass. It is challenging to make a direct comparison in this case as the number of ions surely plays a significant role in the reduction of the mass resolution.

To isolate the effect of ion number on the mass resolution, we measured spectra from pure calcium crystals of varying size. Pure $^{40}$Ca$^+$ crystals of different sizes were loaded into the trap and analysed in the TOF-MS using identical settings. The results of those measurements can be seen in Fig. \ref{fig:fig7}. Here, one can see a decrease in mass resolution as the number of ions increases. This decrease is related to fundamental properties of a TOF-MS; the mass resolution depends strongly on the initial conditions of the ion cloud. \cite{Opsal1985,Schneider2014} A larger Coulomb crystal size results in a larger initial spatial distribution, which causes a decrease in the mass resolution. To further probe the influence of the spatial/energy distribution, we measured the mass resolution with the cooling lasers blocked. This caused a significant increase in the spatial/energy distribution of the ions in the trap. The resulting TOF-MS peak had a significant decreased mass resolution of only R$\sim$40-50. For future reaction experiments, the number of ions in the trap will be kept to $<$ 1000, and thus the mass resolution will be high enough to distinguish any anticipated reaction products.

\subsubsection{\label{sec:results:calibration}Calibration of the TOF-MS}

\begin{figure}[t]
\centering
\includegraphics[width=\columnwidth]{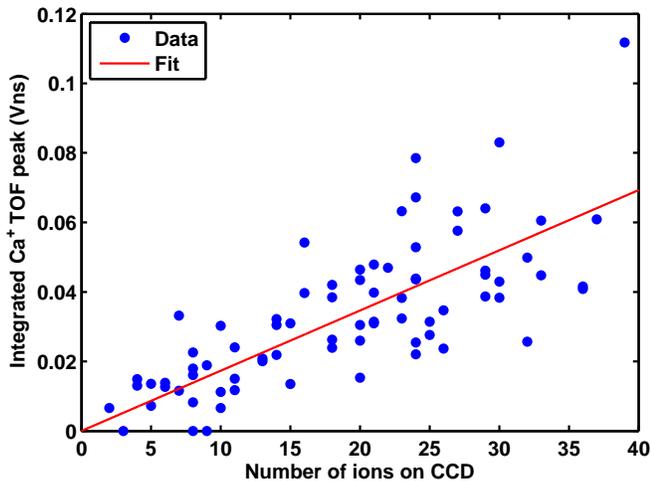}
\caption{Calibration of the MCP signal to ion number. To calibrate the MCP signal, we start by loading small crystals into our trap and counting the number of ions on the image. Then, by extracting those ions into the TOF-MS and integrating the resulting MCP signal, we can determine how the MCP signal relates to the number of ions in the trap. Using this calibration, we can determine the absolute number of ions in the trap for subsequent experiments.}
\label{fig:fig8}
\end{figure}

To fully realize the advantage of using a TOF-MS coupled to an ion trap for studying chemical reactions, mass spectra must be calibrated to allow extraction of absolute numbers of ions present at each mass for a single shot. We use information from the fluorescence images of the crystals to help calibrate our mass spectrometer. The trapped calcium ions represent a deterministic ion source, as the number of ions can be directly counted for small crystals using the fluorescence images. To determine the transfer function of our TOF-MS, we start by loading a small crystal (1 \textendash 40 ions) into the trap and counting the number of ions in the resulting fluorescence image by hand. Then, we eject the ions into the TOF-MS and record a spectrum. We integrate the peak in the spectrum corresponding to $^{40}$Ca$^+$, which produces a number with units of \textsc{\char13}Vns\textsc{\char13}. The results of this procedure are shown in Fig. \ref{fig:fig8}. The plot shows a linear relationship between the number of ions in the trap and the integrated signal from the MCP. The fitted slope of 1.73$\pm$0.07$\times$10$^{-3}$ Vns/ion is then used to determine the number of ions during subsequent reaction experiments. This calibrated ion number is also valid for larger Coulomb crystals, as long as the MCP shows a linear response with the increasing ion numbers, which we have seen for crystals up to at least 1500 ions.  

Using a single trapped ion, we can also measure the probablity of detecting a ion. We extract a single ion into the TOF-MS and observe a signal from the MCP about 38$\pm$3\% of the time. This is the transfer efficiency of the entire system. The largest contribution to the decrease in detection efficiency is the quantum efficiency of the MCP detector, which is 50\textendash 55\%. \cite{Fraser2002}

\subsubsection{\label{sec:results:density}Determining ion density}

\begin{figure}[t]
\centering
\includegraphics[width=\columnwidth]{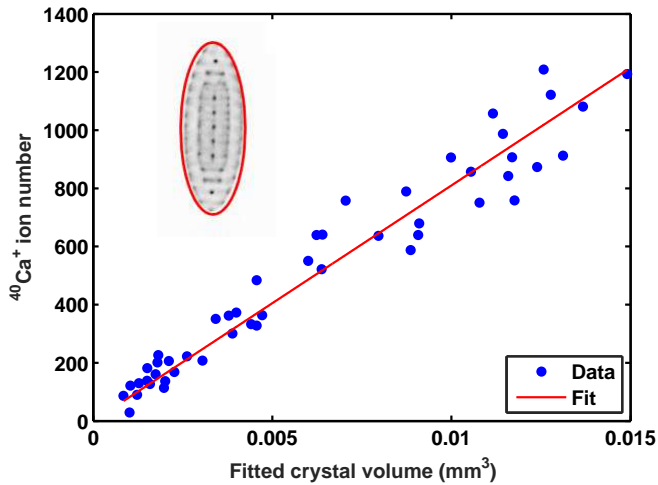}
\caption{Determination of the ion density in our ion trap. We correlate the measured ion number from the TOF-MS to the fitted Coulomb crystal volume taken from the recorded CCD images. From a fit to these data, we can determine the ion density in our trap. The inset shows a Coulomb crystal with an ellipsoidal fit.}
\label{fig:fig9}
\end{figure}

To use the LIT TOF-MS as a platform for studying chemical reactions, one must be able to determine the initial number of trapped ions before the reactions. Although the number of ions loaded into the trap is relatively reproducible, there is some variation and a direct measurement of the initial trap number will help to reduce the uncertainty in the reaction measurements. Since the TOF-MS measurements are destructive, we need to use data from the fluorescence images to determine the initial number of ions. We calculate the number of ions by multiplying the density in the trap by the volume of the Coulomb crystal. The Coulomb crystal volume is determined by an ellipsoidal fit of the CCD image. The  density is then found by fitting a line to the measurements of the integrated MCP signal (calibrated ion number) vs. the volume of the crystal (Fig. \ref{fig:fig9}). From these data, we calculate the ion density for our ion trap to be 7.8$\pm$0.2$\times$10$^7$ ions cm$^{-3}$. This density depends on only the trapping potential and is fixed for any one reaction experiment.
 
\section{\label{sec:summary}Summary and future directions}

We have presented a linear ion trap TOF-MS apparatus for measuring cation-molecule reactions in a controlled environment at reaction temperatures below 1 K.  A linear TOF-MS is coupled radially to an ion trap to enhance identification of the (non-fluorescing) product ions and to accurately determine the number of ions. Our robust and simple design of trap driving electronics addresses the technical challenges of such a system, which include fast quenching of the RF field and application of fast HV pulses to extract ions into a TOF-MS, while achieving a high mass resolution. We achieve a fast quench of the RF fields (within 2 RF cycles) and application of HV pulses with rise times of $<$ 300 ns on the same electrodes. We demonstrate high mass resolution of the detected ions in a wide mass window of $>$ 100 amu within a single shot. Additionally, we have shown an ion detection sensitivity down to a single ion. Finally, we are able to calibrate the mass spectrum to extract absolute ions numbers of all trapped species without the need of molecular dynamics simulations.

Using this apparatus, we investigated the quantum-state controlled reaction between laser cooled $^{40}$Ca$^+$ and room temperature NO radicals.\cite{Greenberg2016} While this measurement showed the strengths of our new experimental set-up, the reaction temperature was above 1 K. In the future, a Stark decelerator\cite{Bethlem1999,Parazzoli2009} will be attached to the ion trap, similar to other proposed experiments.\cite{Willitsch2008,Eberle2015} This will offer enhanced control over the quantum-state population of the neutral molecules, where we can tune the velocity down to 10 m/s with a high energy resolution and nearly single quantum-state selectivity. Then, by introducing state-selective ionization of the trapped molecular cations, true fully quantum-state resolved, cold chemical reactions can be studied in this system.

\begin{acknowledgments}
This work was supported by the National Science Foundation (PHY-1125844 
and CHE-1464997), AFOSR (FA9550-16-1-0117), and NIST.
\end{acknowledgments}

\bibliographystyle{aipnum4-1}

\end{document}